\documentclass[aps,prl,reprint,superscriptaddress]{revtex4-2}

\makeatletter
\def\sectionit{\@startsection
  {section}{1}{0pt}
  {1.5ex plus .2ex minus .2ex}
  {0.8ex}
  {\normalfont\itshape}}
\makeatother

\usepackage[english]{babel}
\usepackage{graphicx}
\graphicspath{{./Images/},{./ImagesAppendix/},{./images/},{./imagesAppendix/}}

\usepackage{amsmath, amssymb, amsfonts}
\usepackage{physics}
\usepackage{bm}
\usepackage{cancel}
\usepackage{braket}
\usepackage{array}
\usepackage{multirow}
\usepackage{float}
\usepackage{commath}
\usepackage{verbatim}
\usepackage{color}
\usepackage{enumitem}
\usepackage{array}   
\usepackage{dsfont}
\usepackage{mathtools}
\usepackage{soul}
\usepackage[colorlinks=true, citecolor=darkBlue, linkcolor=darkBlue, urlcolor=blue]{hyperref}
\usepackage{amssymb}
\usepackage{tabularx}

\definecolor{darkBlue}{rgb}{0.08, 0.13, 0.4}

\usepackage{tikz}
\usetikzlibrary{positioning, arrows.meta}
\usepackage{pgfplots}

\definecolor{THc}{rgb}{0.9,0.3,0.2}

\newcommand{\canc}[1]{}

\newcommand{\orcidicon}[1]{\href{https://orcid.org/#1}{\includegraphics[height=\fontcharht\font`\B]{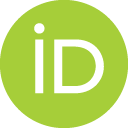}}}

\begin{document}

\title{Phase-dependent role of dissipation across the Aubry-André-Harper transition}
\author{G. Suárez~\orcidicon{0000-0003-1920-8525}}
\affiliation{Institut de Physique Nucléaire, Atomique et de Spectroscopie, CESAM, Université de Liège, 4000 Liège, Belgium}

\author{B. Debecker~\orcidicon{0009-0004-8894-3624}}
\affiliation{Institut de Physique Nucléaire, Atomique et de Spectroscopie, CESAM, Université de Liège, 4000 Liège, Belgium}

\author{F. Cosco\orcidicon{0000-0002-4418-3879}}
\affiliation{Quantum algorithms and software, VTT Technical Research Centre of Finland Ltd, Tietotie 3, 02150 Espoo, Finland}

\author{F. Plastina~\orcidicon{0000-0001-9615-8598}}
\affiliation{Dipartimento di Fisica, Universit\`a della Calabria, 87036 Arcavacata di Rende (CS), Italy}
\affiliation{INFN, Sezione LNF, gruppo collegato di Cosenza}

\author{F. Damanet~\orcidicon{0000-0003-1699-0195}}
\affiliation{Institut de Physique Nucléaire, Atomique et de Spectroscopie, CESAM, Université de Liège, 4000 Liège, Belgium}

\author{N.~Lo~Gullo~\orcidicon{0000-0002-8178-9570}}
\affiliation{Dipartimento di Fisica, Universit\`a della Calabria, 87036 Arcavacata di Rende (CS), Italy}
\affiliation{INFN, Sezione LNF, gruppo collegato di Cosenza}

\author{F.~Perciavalle~\orcidicon{0009-0004-2558-8581}}
\affiliation{Dipartimento di Fisica, Universit\`a della Calabria, 87036 Arcavacata di Rende (CS), Italy}
\affiliation{INFN, Sezione LNF, gruppo collegato di Cosenza}

\date{\today}

\begin{abstract}
We study transport across the Aubry-André-Harper localization transition in the presence of non-Markovian dissipation. For a single particle initially at the center of the chain, we show that bath memory (i.e., finite decay time of bath correlations) plays distinct roles in the two phases. In the extended phase, bath memory qualitatively reshapes the dynamical generator, thereby producing transport patterns that cannot be reduced to a simple rescaling of time. By contrast, in the localized phase, the bath activates motion between localized states and bath memory mainly renormalizes the dynamical timescales. Our results identify localization as a simple filter of non-Markovian effects: memory restructures transport in the extended regime, but survives mainly as a timescale renormalization in the deeply localized regime. 

\end{abstract}

\maketitle

\textbf{\textit{Introduction}} --- Localization captures the failure of quantum transport and is therefore central to condensed matter, many-body physics, and quantum information. Anderson showed that disorder localizes non-interacting particles~\cite{anderson1958absence}, while interactions can lead to many-body localization (MBL) and anomalous diffusion~\cite{alet2018many,pal2010many,abanin2019colloquium,sierant2022challenges,sierant2025many,Settino2020}. Quasiperiodic systems form an especially interesting class of models to study this physics, because localization can emerge without true randomness, as in the Aubry--Andr\'e--Harper (AAH) model~\cite{aubry1980analyticity,harper1955single,schreiber2015observation,modugno2009exponential,modugno2010anderson,roati2008anderson,gottlob2025quasiperiodicity}, Fibonacci chains~\cite{mace2019maby,jiankun2026quantum,strkalj2021many}, and other quasiperiodic lattices~\cite{tabanelli2024reentrant,zhang2025odd,huang1992localization}. Coupling such systems to an environment fundamentally modifies the phenomenology: by exchanging energy with the system, the bath can restore motion, reshape effective hopping processes and partially lift localization~\cite{scocco2024thermalization,longhi2024dephasing,bhakuni2024noise,talia2022logarithmic,turkeshi2022destruction,yang2025dissipation,liu2024dissipation,cui2022localization,xu2025expedited,steiner2026active,kokkinakis2025dephasing,bonca2018dynamics,weidemann2021coexistence}. Dissipation can even produce behavior reminiscent of active matter, sustained by energy exchange between system and environment~\cite{bechinger2016active,vrugt2025exactly,antonov2025engineering,antonov2026modeling,penner2025heat,gipouloux2026active,steiner2026active,yamagishi2024proposal}.

In this context, the role of bath memory remains poorly understood. In the white noise or pure Markovian limit, the environment acts a featureless source of noise. Away from that limit, however, the bath is temporally correlated and spectrally selective. This raises the question of whether memory effects simply change transport timescales or instead qualitatively reshape the generator of the dynamics. Is this effect different in the extended or localized phases?

\begin{figure}[!t]
\centering
\includegraphics[width=0.9\linewidth]{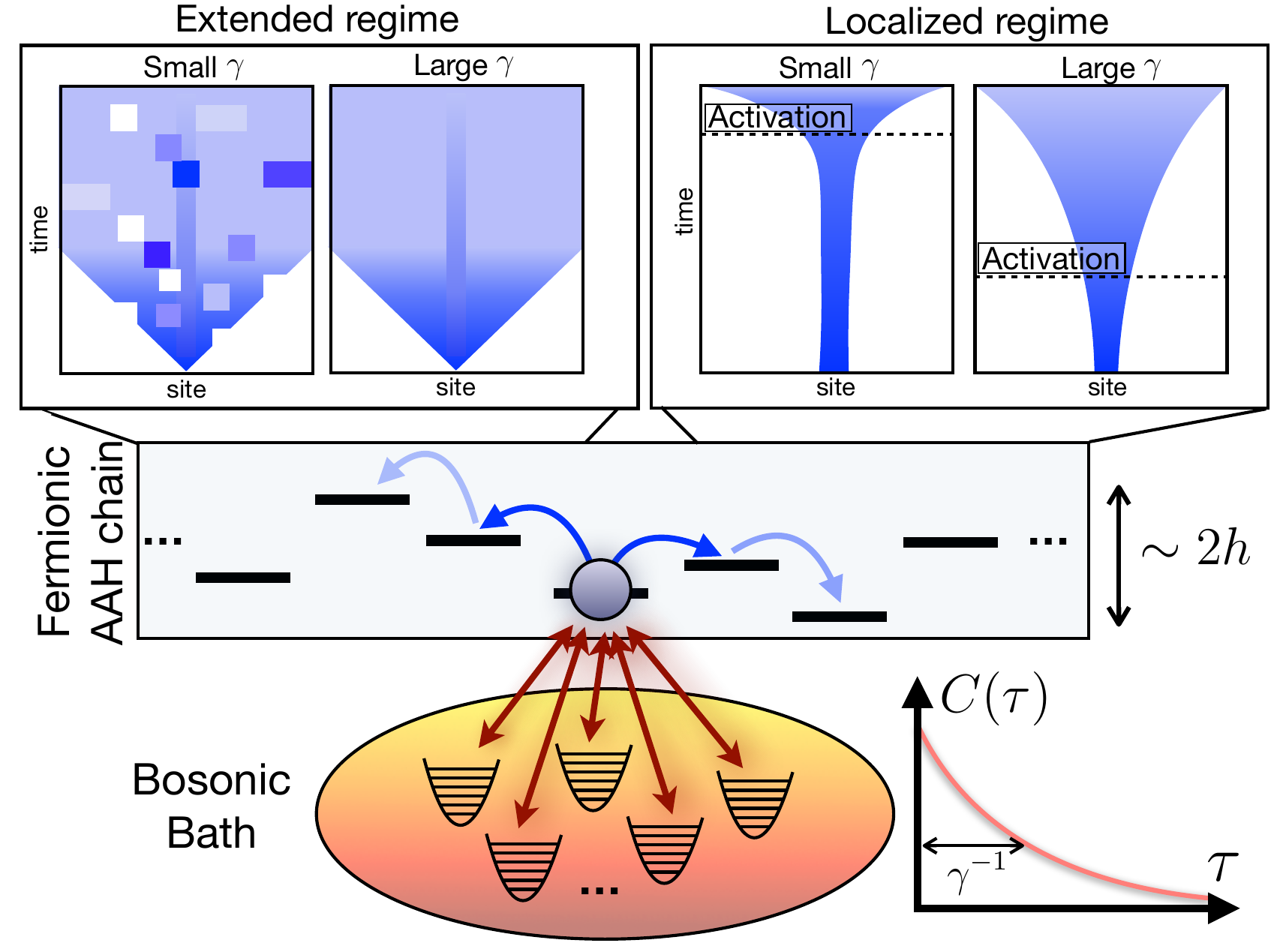}
\caption{Cartoon of the setup: a fermionic quasiperiodic chain is locally coupled at its central site to a bosonic bath, modeled as a non-Markovian environment with exponentially decaying correlations $C(\tau)$ and decay rate $\gamma$. Transport is controlled by system–bath parameters. In the extended phase, $\gamma$ sets the spreading profile, uniform for large $\gamma$ and non-uniform for small $\gamma$. In the localized phase, dissipation induces incoherent hopping, with $\gamma$ mainly fixing the timescale rather than the spatial profile.}
\label{fig:sketch_transport}
\end{figure}
In this Letter, we address this question by elucidating when memory matters for transport in a AAH chain. We couple the central chain to an environment with exponentially decaying correlation function with tunable bath memory~\cite{tanimura2020numerically,jin2008exact,debecker2024controlling}. We probe the dynamics by initializing a single particle at the chain center, accessing the localization–delocalization transition while directly assessing memory effects on transport. Numerical simulations employ the hierarchical equations of motion (HEOM)~\cite{tanimura2020numerically,jin2008exact,lambert2026qutip}, which incorporate bath degrees of freedom into a hierarchy of auxiliary density operators, yielding a closed Liouvillian structure and allowing comparison with standard Markovian descriptions such as Bloch–Redfield (BR) and Lindblad equations~\cite{breuer2007theory,mercurio2025quantum, campaioli2024quantum}.

\begin{figure*}[!t]
\centering
\includegraphics[width=1\linewidth]{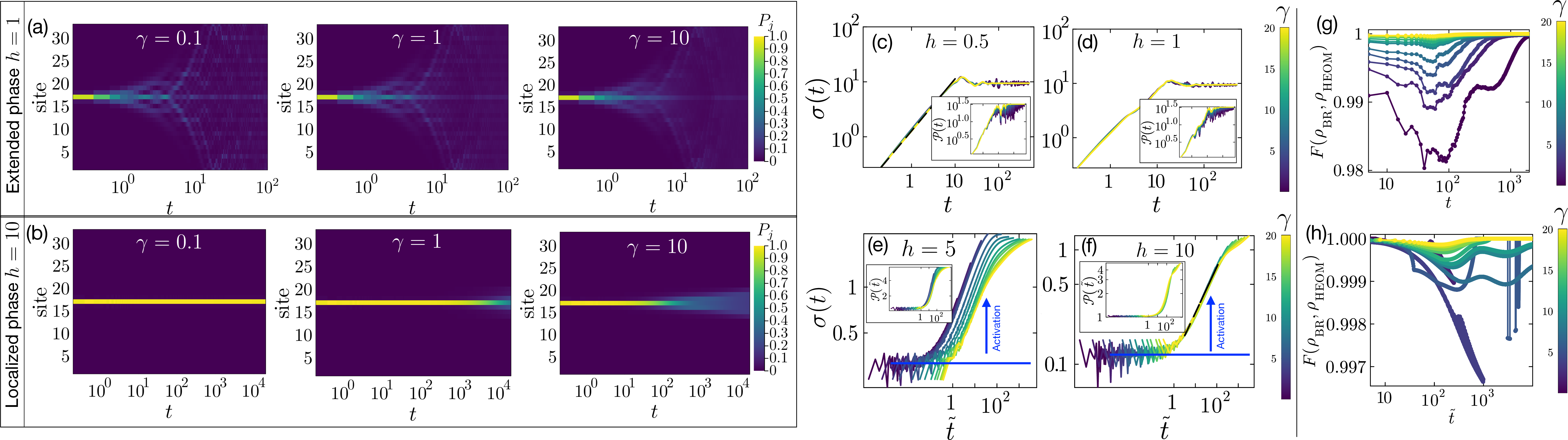}
\caption{Dynamics of a single particle in a dissipative AAH chain across the extended--localized transition. Panels (a) and (b) show the population dynamics in the extended phase ($h=1$) and in the localized phase ($h=10$), respectively, each for three different values of $\gamma$. Panels (c)--(f) display the RMSD, with the PPR shown in the insets, across the phase transition for different values of $h$. In each panel different values of $\gamma$ are considered, as indicated by the color bars. In the localized phase, time is rescaled as $\tilde{t} = \Gamma_{\rm eff}\, t$, where $\Gamma_{\rm eff} = \frac{\kappa \gamma^2}{h^2 + \gamma^2}$. The remaining simulation parameters are $L=33$, $\kappa=2$, and $\mathrm{tier}=6$, different colors correspond to different $\gamma\in [0.1,20]$. The black lines in panels (c) and (f) report the fits of the RMSD for $\gamma=20$, which are respectively $\sigma(t)=1.281\, t^{0.95}$ and $\sigma(t)=0.135 \, t^{0.4}$. Panels (g) and (h) compare the fidelity between the quantum states evolved using HEOM and BR for different values of $\gamma$, with $L=33$ and $\kappa=2$, shown in the extended ($h=1$) and localized ($h=10$) regimes, respectively.}
\label{fig:dynamics}
\end{figure*}
We find that memory affects the dynamical generator differently across phases: in the extended regime, it strongly reshapes the generator, structurally modifying transport; in the localized regime, it primarily renormalizes timescales, producing bath-activated, effectively “active” motion. The dynamics in the localized phase is well captured by a semiclassical master equation describing incoherent, bath-induced hopping with rates tunable via bath properties, demonstrating phase-dependent control of transport in many-body systems, with possible applications to quantum state preparation and control~\cite{damanet2019controlling,li2026macroscopic,begoc2025controlled,chen2025collective,zhao2024dissipative,verstraete2009quantum, tian2026dissipative}. The scheme is pictorially represented in Fig.~\ref{fig:sketch_transport}. 

Non-Markovian environments are hard to control, but superconducting circuits now offer tunable system–reservoir coupling and direct probes of such dynamics~\cite{zhang2025experimental}, while recent trapped-ion experiments demonstrate an impressive control in reservoir-engineering~\cite{So2025Q, So2024E, Sun2025Q}. Building on insights into thermal effects in localization from ultracold atoms~\cite{lonard2023probing}, a key perspective is to study how memory, disorder, and temperature jointly govern dynamics across the extended–localized transition within superconducting circuits.

\textbf{\textit{Model $\&$ methods }} --- We consider a system composed of a spinless fermionic quantum chain of $L$ sites, described by the Aubry-André-Harper (AAH) model~\cite{aubry1980analyticity,harper1955single,modugno2009exponential, modugno2010anderson, roati2008anderson}:
\begin{equation}
\hat{H}_f^{(\mathrm{AAH})} = 
J \sum_{j=1}^{L-1} \Bigl( \hat{c}_j^\dagger \hat{c}_{j+1} + \text{H.c.} \Bigr)
+ \sum_{j=1}^L V_j \hat{n}_j,
\end{equation}
where $\hat{c}_j$ ($\hat{c}_j^\dagger$) are fermionic annihilation (creation) operators, 
$\hat{n}_j = \hat{c}_j^\dagger \hat{c}_j$ is the number operator, 
and $V_j = h \cos(2\pi \beta j + \phi)$ is the AAH quasiperiodic potential. Here, $\beta = (\sqrt{5}-1)/2$ is the inverse golden ratio, 
and $\phi$ is an arbitrary phase; in the remainder of the paper we set $\phi = 0$. The model exhibits a phase transition from an extended phase to a localized phase at the critical value $h_c = 2J$. For $h < h_c$ the system is in the extended phase, while for $h > h_c$ it enters the localized phase. Overall, the AAH model enables deterministic control over localization properties. In the following, we set $J=1$, such that all energies and times are measured in units of $J$ and $J^{-1}$, respectively.

The system is coupled to an environment through its central site $j_0$, and the full Hamiltonian reads
\begin{equation}
    \hat{H} = \hat{H}_f^{(\mathrm{AAH})} + \sum_k \omega_k \hat{b}_k^\dagger \hat{b}_k 
    + \hat{n}_{j_0} \otimes \underbrace{\sum_k \lambda_k \left( \hat{b}_k^\dagger + \hat{b}_k \right)}_{\hat{B}}.
\end{equation}
The bath is modeled as a collection of independent harmonic oscillators, where $\hat{b}_k$ ($\hat{b}_k^\dagger$) denote the bosonic annihilation (creation) operators of the $k$-th mode and $\omega_k$ its frequency, $\hat{B}$ is the operator of the bath that couples it to the system. The parameter $\lambda_k$ represents the coupling strength between each bath oscillator and the fermionic chain. We consider a chain with an odd number of sites $L$, such that the central site is $j_0 = (L+1)/2$. The full setup is pictorially represented in Fig.~\ref{fig:sketch_transport}.

We are interested in studying the dynamics of the fermionic chain which, in this setting, constitutes an open non-Markovian quantum system. It is convenient to introduce the spectral density $J(\omega) = 2\pi \sum_k |\lambda_k|^2 \delta(\omega - \omega_k)$, and the bath correlation function $C(t)=\braket{\hat{B}(t)\hat{B}(0)}$ that depends on $J(\omega)$ and the temperature of the bath.

We compute the reduced dynamics of the fermionic system using the hierarchical equations of motion (HEOM) approach~\cite{tanimura2020numerically,jin2008exact,debecker2024controlling,debecker2024spectral,debecker2025role,lambert2026qutip,HierarchicalEOM.jl2023,bai2024hierarchical,cirio2025hierarchical,lambert2023qutip}. The key idea of HEOM is to treat non-Markovian open-system dynamics by embedding the system into an enlarged space. Specifically, the physical density matrix is incorporated into a vector $|\rho\rangle\!\rangle$ that includes, in addition, a hierarchy of auxiliary density operators. These auxiliary operators encode the environment and its memory effects, so that the influence of the bath is captured dynamically rather than through a Markovian approximation. The reduced state of the system is obtained by projecting onto (or equivalently tracing out) the auxiliary components. The full set of operators evolves in Liouville space under a superoperator $\mathcal{L}_{\mathrm{HEOM}}$, whose spectrum determines the relaxation timescales of the dynamics~\cite{debecker2024controlling}: $
\frac{d |\rho\rangle\!\rangle}{dt}=\mathcal{L}_{\rm HEOM}|\rho\rangle\!\rangle$.

The structure of $\mathcal{L}_{\mathrm{HEOM}}$ is fixed by expressing the bath correlation function as a sum of exponentials, which generates the hierarchy of auxiliary operators. In practice, the hierarchy is truncated at a finite tier, becoming exact in the infinite-tier limit. We consider the bath correlation function $C(t) = \frac{\kappa \gamma}{2} e^{-\gamma |t|}$, where $\kappa$ sets the total weight of the correlations, with $\int_0^\infty C(t)\, dt = \frac{\kappa}{2}$, and $\gamma$ controls the timescale of the environmental correlations. In the limit $\gamma \to \infty$, the correlation function approaches a delta function, $C(t) \to \kappa/2 \, \delta(t)$, recovering the Lindblad limit. For finite $\gamma$, the dynamics depart from this limit due to the introduction of memory effects arising from the finite temporal extent of the bath correlations.

\textbf{\textit{Dynamics across the phase transition and role of bath correlations}} --- We study the dynamics of a particle initially localized at site $j_0$, described by the state $\ket{\psi(0)} = \ket{j_0}$. As the system evolves in time, the particle spreads over the lattice, and its probability of occupying site $j$ at time $t$ is given by $P_j(t) = \operatorname{Tr}[\rho(t)\hat{n}_j]$. To quantify spreading, we use the root-mean-square displacement (RMSD) $\sigma(t)=\sqrt{\sum_j (j-j_0)^2 P_j(t)}$, which measures the average distance from the initial site $j_0$. Its scaling $\sigma(t)\sim t^\alpha$ characterizes transport: $\alpha=1$ (ballistic), $\alpha=1/2$ (diffusive), $\alpha<1/2$ (subdiffusive), and $1/2<\alpha<1$ (superdiffusive/sub-ballistic). However, the RMSD alone cannot distinguish between genuine delocalization and the mere spreading of a localized or structured wave packet. For instance, $\sigma(t)$ can grow due to the motion of a few distant tails or peaks, even if most of the population remains confined. To resolve this ambiguity, we introduce the population participation ratio (PPR) $\mathcal{P}(t)=(\sum_j P_j(t)^2)^{-1}$, which measures the effective number of significantly occupied sites. Unlike the RMSD, the PPR is sensitive to how uniformly the population is distributed: it remains small if the dynamics is confined to a few sites (even if those sites move), and grows only when the population spreads broadly across the system. Therefore, combining $\sigma(t)$ and $\mathcal{P}(t)$ allows us to discriminate between true delocalization and compact or structured transport.

Before analyzing the RMSD and PPR, we first report the population dynamics in Fig.~\ref{fig:dynamics}(a) and (b), highlighting the effect of the dissipation factor $\gamma$ on the particle evolution in both the extended and localized phases. In the extended phase, $\gamma$ mainly affects the spatial distribution without significantly changing the overall timescale. For large $\gamma$ (Lindblad regime), the population remains smooth, while for small $\gamma$ it becomes fragmented due to coherent interference and less homogeneous spreading. In the localized phase, by contrast, dissipation breaks localization by enabling bath-assisted motion, leading to an effective active transport. Here, increasing $\gamma$ primarily rescales the dynamics in time, with little impact on the spatial profile, unlike in the extended phase, where $\gamma$ qualitatively reshapes the distribution.

The dynamics of $\sigma(t)$ and $\mathcal{P}(t)$ provides a clear characterization of the impact of dissipation on the system's evolution. We first discuss the extended phase, shown in Figs.~\ref{fig:dynamics}(c),(d). In the absence of dissipation and external potentials, transport is ballistic, $\sigma(t)\sim t$, until the wavepacket reaches the system boundaries, after which reflections induce oscillations around a quasi-stationary value. A weak AAH potential slightly reduces the ballistic growth without changing its character. Dissipation mainly damps post-reflection oscillations, leading to smoother relaxation to a steady profile. Far from the Lindblad regime, this damping is counteracted and oscillations are weakly restored, especially at small $h$. Overall, the RMSD is only weakly sensitive to $\gamma$. In contrast, the effect of $\gamma$ is clearly captured by the PPR, which exhibits a more pronounced oscillatory behavior for smaller values of $\gamma$. These results confirm that the factor $\gamma$ has a structural impact on the dynamics in the extended regime.

In the localized phase, transport is activated by the bath, as shown in Fig.~\ref{fig:dynamics}(b). To understand the mechanism underlying this activation, we develop a semiclassical description (see SM~\cite{suppmat}), following Ref.~\cite{bhakuni2024noise}. We assume the system is described by
$\hat{H}(t)=\hat{H}_f+\sqrt{g/2}\,\eta(t)\,\hat{n}_{j_0}$,
with autocorrelation $\overline{\eta(t)\eta(t')}=\gamma e^{-\gamma |t-t'|}$. For $g \propto \kappa$, this driving induces incoherent transitions between localized eigenstates, leading to a rate equation
$\partial_t P_m=\sum_n (\Gamma_{mn}P_n-\Gamma_{nm}P_m)$,
written in the Hamiltonian eigenstates. The transition rates take the form
$\Gamma_{nm}\sim e^{-2(|j_n-j_0|+|j_m-j_0|)/\xi}\,L(\epsilon_n-\epsilon_m)$,
where $\xi$ is the localization length and $L(\omega)=\frac{\kappa \gamma^2}{\omega^2+\gamma^2}$ is a Lorentzian distribution. In the strongly localized regime $h>h_c$, the dynamics reduces to incoherent hopping, with generally asymmetric rates $\Gamma_{j,j+1}\neq \Gamma_{j,j-1}$. A hop between neighboring sites requires exchanging an energy $\epsilon_j-\epsilon_{j\pm1}\sim h$ with the bath, so the corresponding rate is set by $L(h)$,
$\Gamma_{\rm eff}=L(h)=\frac{\kappa \gamma^2}{h^2+\gamma^2}$. For small $\gamma$, $L(\omega)$ is sharply peaked at $\omega=0$, strongly suppressing transitions involving finite energy exchange $h$. This suppression originates from the finite decay time $\gamma^{-1}$ of the bath, which limits its ability to resolve and supply the energy required for hopping, leading to slow, bath-activated transport in the localized phase.

Figs.~\ref{fig:dynamics}(e),(f) show the dynamics of the RMSD and PPR in the localized phase, with time rescaled as $\tilde{t} = \Gamma_{\rm eff} t$. Dissipation activates motion in this regime. Bath-activated transport is effectively $\gamma$-independent: the RMSD and PPR curves collapse when time is rescaled by $\Gamma_{\rm eff}$, indicating bath-activated motion occurs, with $\gamma$ mainly setting transport timescales. Other dissipation parameters, such the the oscillation frequency of the bath-correlation function, can instead produce qualitative changes in the RMSD dynamics~\cite{suppmat}.

The transport shows an initial quasi-stationary regime ending at $\tilde{t}\approx 1$ ($t\approx \Gamma_{\rm eff}^{-1}$), corresponding to the inverse rate for the particle to exchange energy $h$ with the environment and initiate motion. This leads to bath-activated dynamics followed by subdiffusive growth, $\sigma(t) \approx D \, \tilde{t}^{\alpha} = \left[D \, \Gamma_{\rm eff}^{\alpha}\right] t^{\alpha}$, where $D$ is a subdiffusive prefactor independent of $\gamma$ and $\kappa$, but dependent on $h$ and on the localization properties of the AAH eigenmodes~\cite{suppmat}. For $h=10$, the dynamical exponent is $\alpha\approx 0.4$. At later times, transport slows further to an even slower subdiffusive regime with $\alpha\approx 0.15$. In this regime the transport is predominantly incoherent, indeed the system evolves such that its state can be represented as a diagonal density matrix in the localized basis: 
$\rho(t) \approx \sum_{j \in \ell(t)} P_j(t) \, \ket{j}\bra{j}$, where $P_j(t)$ gives the occupation probability of site $j$ and $\ell(t)$ denotes the set of dynamically relevant sites~\cite{suppmat}.

Finally, in Fig.~\ref{fig:dynamics}(g),(h), we compare HEOM and Markovian BR dynamics, by studying the fidelity between the time-evolved density matrices obtained with the two methods. We can observe that regardless of the value of $\gamma$, the dynamics of this system is captured by a second order process as evidenced by the extremely high fidelity of the BR approach. Notice that the drop in fidelity is tiny in the localized phase, and it is only noticeable in the transient regime, for the extended phase and low values of $\gamma$. This is  expected as the BR equation is Markovian ~\cite{breuer2007theory,campaioli2024quantum}. The study of the fidelity  agrees with the analysis of the spectrum in ~\cite{suppmat}, where appreciable difference between the different approaches only exists for the extended phase.  

\begin{figure}[!t]
\centering
\includegraphics[width=\linewidth]{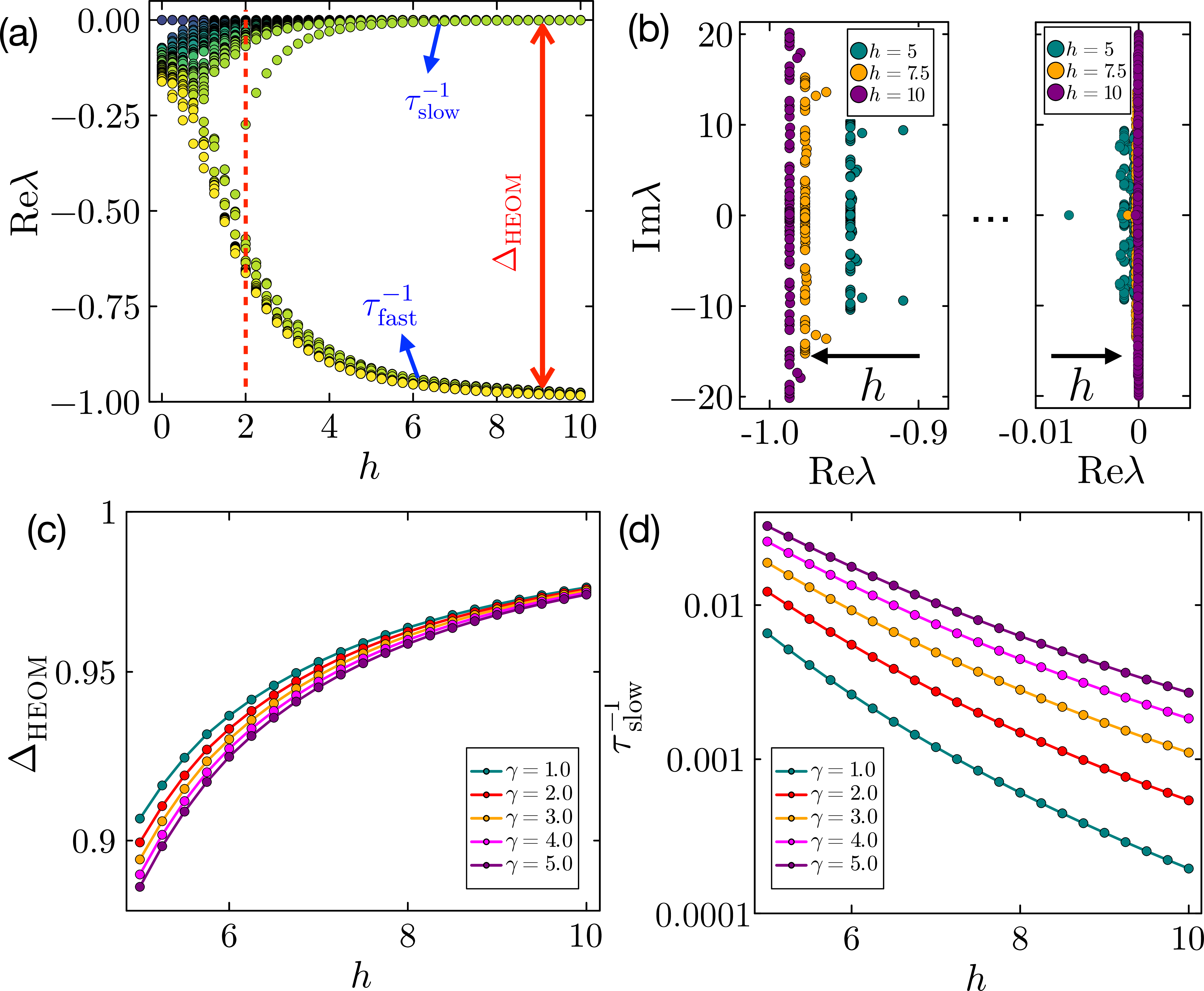}
\caption{Features of the HEOM spectrum across the extended-to-localized transition.
(a) Real parts of $\mathcal{L}_{\rm HEOM}$ eigenvalues, showing clustering that identifies $\Delta_{\rm HEOM}$, $\tau_{\rm fast}^{-1}$, and $\tau_{\rm slow}^{-1}$. Parameters: $L=15$, $\gamma=2$, tier$=6$. (b) Real and imaginary parts of $\mathcal{L}_{\rm HEOM}$ eigenvalues in the localized phase for different $h$ and $L=33$, zoomed on fast and slow clusters. Parameters: $\gamma=2$, tier$=6$. For both panels (a) and (b), we consider the $L^2$ eigenvalues of $\mathcal{L}_{\mathrm{HEOM}}$ with the largest real parts. (c),(d) $\Delta_{\rm HEOM}$ and $\tau_{\rm slow}^{-1}$ vs $h$ in the localized phase for various $\gamma$. $L=15$; tier$=8$ ($\gamma=1$), tier$=6$ ($\gamma=2,3$), tier$=5$ ($\gamma=4,5$).
In all panels, $\kappa=2$.}
\label{fig:HEOM_spectrum}
\end{figure}

\textbf{\textit{HEOM spectrum across the phase transition}} ---
\label{sec:HEOM_spectrum}
To support results on the dynamics, we analyze the HEOM spectrum through the eigenvalues of $\mathcal{L}_{\rm HEOM}$, with particular focus on their real parts, which govern the relaxation dynamics. We first analyze the HEOM spectrum across the extended-to-localized phase transition for a set of fixed parameters, see Fig.~\ref{fig:HEOM_spectrum}(a). 
As the disorder strength $h$ increases, the eigenvalues begin to cluster into two distinct groups: one associated with short timescales, $\tau_{\rm fast}$, and another governing long timescales, $\tau_{\rm slow}$. The two clusters are separated by a gap, $\Delta_{\rm HEOM}$, which emerges as a feature of the localized regime. 
Here, $\tau_{\rm fast}$ and $\tau_{\rm slow}$ are identified, respectively, as the minimum and maximum $|\mathrm{Re}\lambda|^{-1}$ within each cluster. This gapped structure in the generator spectrum indicates metastability~\cite{macieszczak2016towards,macieszczak2021theory,shang2026steady}, with ultra-slow relaxation expected at intermediate times, $\tau_{\rm fast} \ll t \ll \tau_{\rm slow}$. The timescale $\tau_{\rm slow}$ plays a direct role in the dynamics: the crossover between the two subdiffusive regimes observed in Fig.~\ref{fig:dynamics}(f) occurs approximately at $\tilde{\tau}_{\rm slow} = \Gamma_{\rm eff}\tau_{\rm slow}$~\cite{suppmat}. This corresponds to the point at which the slow eigenvalue cluster of the dynamical generator becomes dominant, marking the end of the metastable regime mentioned above.

Fig.~\ref{fig:HEOM_spectrum}(b) shows the real and imaginary parts of the $\mathcal{L}_{\rm HEOM}$ eigenvalues for the two clusters of states at different values of $h$ with $L=33$ in the localized phase. We observe that, as $h$ increases, the eigenvalues tend to collapse onto two lines in the $(\mathrm{Re}\lambda,\mathrm{Im}\lambda)$ plane. The imaginary parts span a region proportional to the bandwidth of the closed fermionic system, while the real parts define the $\Delta_{\rm HEOM}$ gap. When $h$ is large enough, the imaginary parts of the eigenvalues coincide the Bohr frequencies $\lambda \approx -i (\epsilon_n - \epsilon_m)$ with $\hat{H}_f \ket{n} = \epsilon_n \ket{n}$.

Subsequently, we comment on the behavior of $\Delta_{\rm HEOM}$ and $\tau_{\rm slow}^{-1}$ as functions of $h$ in the localized phase, shown in Fig.~\ref{fig:HEOM_spectrum}(c),(d). We compare results for different values of $\gamma$. The gap $\Delta_{\rm HEOM}$ exhibits a similar qualitative trend for all choices of $\gamma$, increasing as $h$ grows, and converging towards $\kappa/2$ (see SM~\cite{suppmat}). In parallel, we analyze $\tau_{\rm slow}^{-1}$ [panel (c)]. As with the gap, its qualitative dependence on $h$ is largely independent of $\gamma$, while the quantitative values differ significantly. For example, at $h=10$, we find $\tau_{\rm slow}\,(\gamma=1)$ exceeds $\tau_{\rm slow}\,(\gamma=5)$ of more than one order of magnitude. Overall, while varying $\gamma$ does not alter the qualitative structure of the HEOM spectrum, it has a pronounced impact on the associated dynamical timescales.

Finally, in the SM~\cite{suppmat}, we compare the spectrum of dynamical generators obtained using the HEOM method with those from Markovian approaches such as Bloch–Redfield and Lindblad. As expected, the results agree in the large-$\gamma$ limit. For small $\gamma$, the spectra in the localized phase differ only by a rescaling, whereas in the extended phase their structures are qualitatively different, indicating that the nature of dissipation has a profound impact on the system’s dynamics.

\textbf{\textit{Conclusions and outlook}}--- We studied the dynamics of a fermionic particle in a quasiperiodic potential locally coupled to a bosonic bath, forming an open non-Markovian system. We showed that bath-correlation features produce phase-dependent effects: in the extended phase, they qualitatively modify transport. In the localized phase, dissipation induces incoherent, bath-assisted transport, yielding effective “active” motion absent in the closed system, with the bath correlation function that primarily controls the dynamical timescales. In this way, dissipation restores transport by breaking perfect localization, unlike in quantum Zeno settings and measurement-induced phase transitions, where dissipation instead suppresses motion and induces localization~\cite{turkeshi2021measurement,maniscalco2008protecting,piccitto2024impact, jin2024measurement}.

Using the HEOM approach and comparison with Markovian descriptions, we show that the spectrum of the dynamical generator distinguishes extended and localized phases. In the extended phase, the bath correlation features mainly reshape the generator, modifying transport patterns without strongly affecting timescales. In the localized phase, the spectrum forms a strongly clustered structure, where the slowest-relaxation cluster $\tau_{\rm slow}$ governs long-time dynamics and transport becomes subdiffusive, with an additional slower regime due to exponentially suppressed long-range hopping. 
An effective semiclassical description confirms that the bath sets the dominant rate, so rescaling time collapses RMSD curves, indicating that, in the localized limit, dissipation primarily controls timescales without altering spreading behavior.

These results demonstrate that bath-correlation features govern transport in open quantum systems and admit an effective description in terms of incoherent, bath-activated hopping rates. More generally, this framework enables systematic control of transport via bath engineering and the design of dissipation-driven, dissipation-controlled active-motion protocols in many-body systems, with broad relevance to condensed matter, quantum optics, and quantum information~\cite{li2026macroscopic,schonleber2015quantum,kitson2024rydberg,damanet2019controlling,gou2020tubanle,pellitteri2026exact,stanzione2025tailoring,weidemann2021coexistence}.

\textbf{\textit{Acknowledgements}}--  The HEOM simulations are performed by using the package \texttt{HierarchicalEOM.jl}~\cite{HierarchicalEOM.jl2023,QuantumToolbox.jl2025}. FP, NLG and FP acknowledge A. Palamara for fruitful discussions. This work was partially funded by the PNRR MUR Project No. PE0000023-NQSTI through the
secondary project ThAnQ. GS was supported by the University of Liege under Special Funds for
Research, IPD-STEMA Programme.

\clearpage
\widetext
\begin{center}
\textbf{\large Supplemental Material for ``Phase-dependent role of dissipation across the Aubry-André-Harper transition''}
\end{center}

\setcounter{equation}{0}
\setcounter{figure}{0}
\setcounter{table}{0}

\renewcommand{\thefigure}{S\arabic{figure}}
\renewcommand{\theequation}{S\arabic{equation}}

\section{Detailed analysis of the deeply localized regime}
\begin{figure}[!h]
\centering
\includegraphics[width=0.85\linewidth]{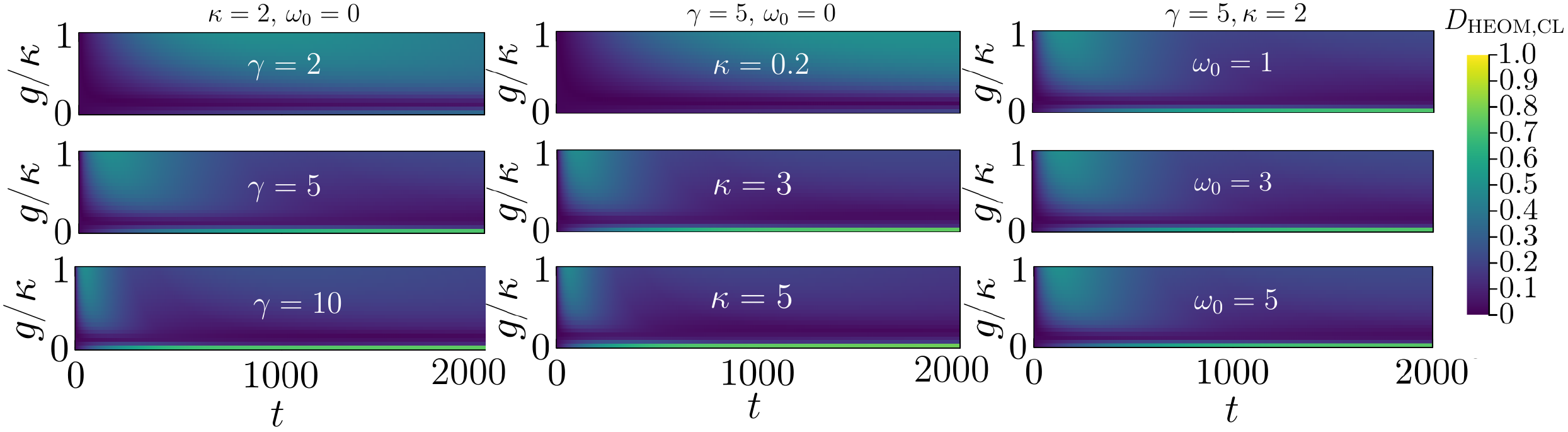}
\caption{Distance between time-evolved populations obtained with HEOM and semiclassical approach. Different dissipation parameters are considered and $L=33, \,h=10$. The HEOM simulations are performed by using tier$\,=6$.}
\label{fig:comparison_heom_semi}
\end{figure}
In this section, we analyze the dynamics in the deeply localized regime. We first derive a semiclassical master equation that captures incoherent, bath-activated hopping along the chain. We then compare its predictions with HEOM simulations to assess accuracy. Next, we demonstrate the collapse of the RMSD dynamics under time rescaling, highlighting the role of bath-induced rates. Finally, we elucidate the fundamentally incoherent nature of particle transport in this regime.
\subsection{Derivation of the semiclassical master equation}
Let us consider a system governed by the time-dependent Hamiltonian
\begin{equation}
    \hat{H}(t) = \hat{H}_f + \hat{V}(t), \quad 
    \hat{V}(t) = \sqrt{\frac{g}{2}}\,\eta(t)\,|j_0\rangle\langle j_0|.
\end{equation}
Here, the system is subject to temporally correlated noise $\eta(t)$, whose correlations decay exponentially in time, 
\begin{equation}
    \overline{\eta(t)\eta(t')} \equiv \mathcal{Q}(t-t') = \textrm{Re} \left[ \gamma\, e^{-\gamma |t-t'|} e^{-i\omega_0 (t-t')}\right]=\gamma\, e^{-\gamma |t-t'|} \cos(\omega_0 (t-t')).
\end{equation}
Such correlations introduce memory effects that significantly influence the dynamics. Thus, we follow the approach presented in Ref.~\cite{bhakuni2024noise} in the presence of correlated noise. This correlation function extends the form used in the main text to include an additional oscillatory behavior.

We take $\hat{H}_f$ to be the Aubry-Andr\'e Hamiltonian in the localized phase, which can be diagonalized as 
\begin{equation}
    \hat{H}_f = \sum_n \epsilon_n\,|n\rangle\langle n|, \quad |n\rangle = \sum_j \phi_n(j)\,|j\rangle,
\end{equation}
with exponentially localized eigenstates 
\begin{equation}
    \phi_n(j) = \frac{e^{-|j-j_n|/\xi}}{\mathcal{N}}, \quad \xi = \frac{1}{\ln(h/2J)},
\end{equation}
where $j_n$ denotes the site around which the state $|n\rangle$ is localized, $\mathcal{N}$ is a normalization factor.

\begin{figure}[!t]
\centering
\includegraphics[width=0.55\linewidth]{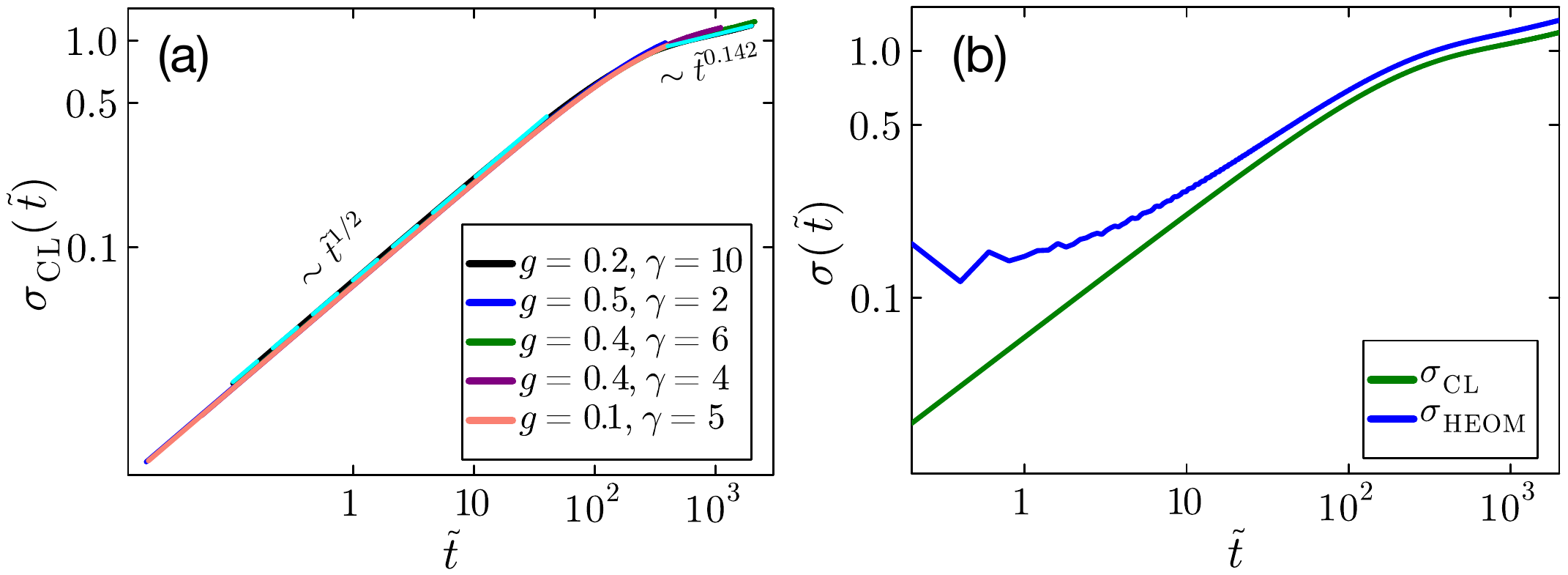}
\caption{Semiclassical dynamics and comparison with HEOM results. Panel (a) shows the time evolution of the RMSD, $\sigma_{\rm CL}$, obtained within the semiclassical approach for different values of $g$ and $\gamma$, $L = 101$, and $h = 10$. Panel (b) compares $\sigma_{\rm CL}$ with the HEOM result $\sigma_{\rm HEOM}$ for $g = 0.2$, $\gamma = 10$ (semiclassical), and $\kappa = 2$, $\gamma = 10$ (HEOM). All other parameters are the same as in panel (a), namely $L = 101$, and $h = 10$, $\omega_0=0$ is fixed. The HEOM calculations are performed with a truncation tier of $6$.}
\label{fig:semicalssical_dyn}
\end{figure}
The time evolution of the state $|\psi(t)\rangle$ in the eigenbasis $|n\rangle$ of $\hat{H}_f$ can be written as
\begin{equation}
\frac{d c_n}{dt} = -i \sum_{n'}\langle n | \hat{V}(t) | n' \rangle \, c_{n'}(t) \, e^{-i (\epsilon_{n'} - \epsilon_n) t},
\end{equation}
In integral form, this corresponds to
\begin{equation}
c_n(t) = c_n(0) - i \int_0^t dt' \sum_{n'} \langle n | \hat{V}(t') | n' \rangle \, c_{n'}(t') \, e^{-i (\epsilon_{n'} - \epsilon_n) t'}.
\end{equation}
Substituting the LHS into the RHS and ignoring terms of order $\mathcal{O}(V^2)$, we obtain
\begin{equation}
c_n(t) \approx c_n(0) - i \int_0^t dt' \sum_{n'} \langle n | \hat{V}(t') | n' \rangle \, c_{n'}(0) \, e^{-i (\epsilon_{n'} - \epsilon_n) t'}.
\end{equation}
We now assume that the system is initialized in state $\ket{m}$, $c_{n'}(0) = \delta_{n' m}$, we have
\begin{equation}
c_n(t) = \delta_{nm} - i \int_0^t dt' \langle n | \hat{V}(t') | m \rangle \, e^{-i (\epsilon_m - \epsilon_n) t'}.
\end{equation}
Thus, the transition probability from $\ket{m}$ to $\ket{n}$ is
\begin{equation}
|c_n(t)|^2 
\approx \int_0^t dt' \int_0^t dt'' \, 
\langle n | \hat{V}(t') | m \rangle \langle m | \hat{V}(t'') | n \rangle \, 
e^{-i (\epsilon_m - \epsilon_n)(t' - t'')}.
\end{equation}
After averaging over the noise, the transition probability reads
\begin{equation}
|c_n(t)|^2 
\approx |\phi_n(j_0)|^2 |\phi_m(j_0)|^2 \int_0^t dt' \int_0^t dt'' \, \gamma \frac{g}{2}\, e^{-\gamma |t' - t''|}\cos(\omega_0 (t' - t''))
e^{-i (\epsilon_m - \epsilon_n)(t' - t'')}\equiv |\phi_n(j_0)|^2 |\phi_m(j_0)|^2 \gamma \frac{g}{2} I(t),  
\end{equation}
where we have introduced the integral
\begin{equation}
    I(t) = \frac{1}{2}\underbrace{\int_0^t \int_0^t dt'\, dt'' e^{-\gamma |t' - t''|} e^{i \Delta_{nm}^{(+)}(t' - t'')}}_{\equiv J_{\Delta_{nm}^{(+)}}(t)} + \underbrace{\frac{1}{2}\int_0^t \int_0^t dt'\, dt'' e^{-\gamma |t' - t''|} e^{i \Delta_{nm}^{(-)}(t' - t'')}}_{\equiv J_{\Delta_{nm}^{(-)}}(t) }, \quad \Delta_{nm}^{(\pm)} \equiv (\epsilon_n - \epsilon_m)\pm \omega_0.
\end{equation}
The integral $J_\Delta(t)$ can be evaluated explicitly, yielding
\begin{equation}
    J_\Delta(t)= \frac{2\gamma t}{\gamma^2 + \Delta^2} + 2\frac{(\Delta^2 - \gamma^2)}{(\gamma^2 + \Delta^2)^2} - 2\frac{(\Delta^2 - \gamma^2)e^{-\gamma t}\cos(\Delta t)} {(\gamma^2 + \Delta^2)^2} - \frac{4\Delta\gamma e^{-\gamma t}\sin(\Delta t)}{(\gamma^2 + \Delta^2)^2}.
\end{equation}
\begin{figure}[!t]
\centering
\includegraphics[width=0.8\linewidth]{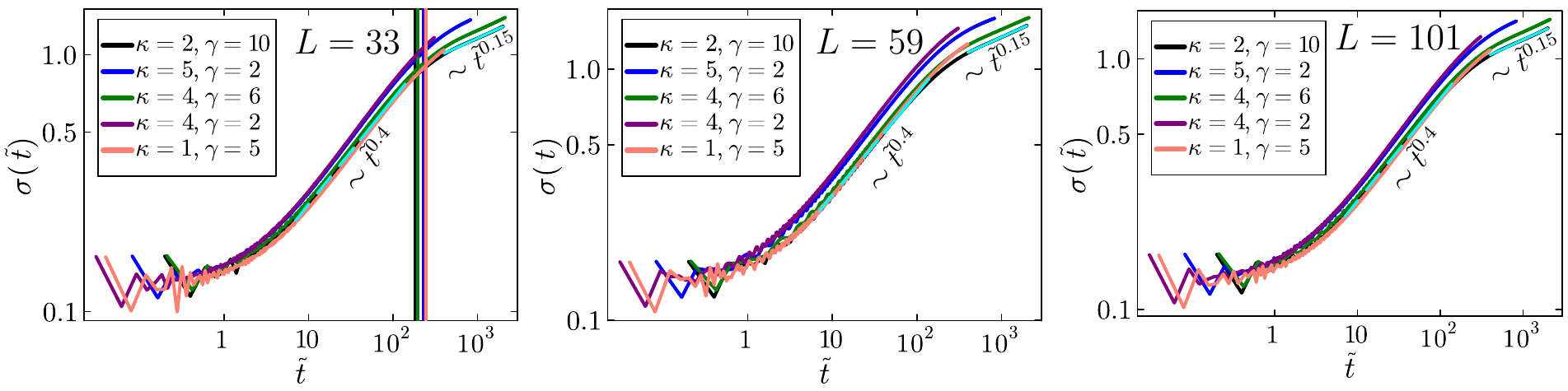}
\caption{RMSD collapse in the localized phase ($h$ = 10) via time rescaling $\tilde{t}=\Gamma_{\rm eff} t$, for $L = 33, 59, 101$ (separate panels) and different dissipation parameters at fixed $\omega_0=0$. The vertical lines in the first panel indicate $\tilde{\tau}_{\rm slow}=\Gamma_{\rm eff} \tau_{\rm slow}$ obtained from the HEOM spectrum for each choice of the parameters. Cyan lines show short- and long-time fits (tier $\,= 6$).}
\label{fig:collapse}
\end{figure}
For $\gamma$ sufficiently large, so that $t \gg 1/\gamma$ is quickly reached, the integral is dominated by the first term, giving $J_{\Delta}(t) \approx \frac{2 \gamma t}{\gamma^2 + \Delta^2}$ and so $I(t)\approx \frac{\gamma t}{\gamma^2 + \left(\Delta_{nm}^{(+)}\right)^2} + \frac{ \gamma t}{\gamma^2 + \left(\Delta_{nm}^{(-)}\right)^2} $, which implies
\begin{align}
    |c_n(t)|^2 \approx &  |\phi_n(j_0)|^2 |\phi_m(j_0)|^2 \frac{\gamma g}{2}\left( \frac{\gamma t}{\gamma^2 + \left(\Delta_{nm}^{(+)}\right)^2} + \frac{ \gamma t}{\gamma^2 + \left(\Delta_{nm}^{(-)}\right)^2}\right)=\\&\frac{e^{-2|j_m-j_0|/\xi} e^{-2|j_n-j_0|/\xi}}{2\mathcal{N}^2} \left( \frac{g\gamma^2 t}{\gamma^2 + \left(\Delta_{nm}^{(+)}\right)^2} + \frac{ g\gamma^2  t}{\gamma^2 + \left(\Delta_{nm}^{(-)}\right)^2}\right),
\end{align}
which means that the transition rate from the state $\ket{m}$ to the state $\ket{n}$ reads \cite{bhakuni2024noise}
\begin{equation}
    \Gamma_{nm}=\frac{e^{-2|j_m-j_0|/\xi} e^{-2|j_n-j_0|/\xi}}{2\mathcal{N}^2} \left( \frac{g\gamma^2}{\gamma^2 + \left(\Delta_{nm}^{(+)}\right)^2} + \frac{ g\gamma^2 }{\gamma^2 + \left(\Delta_{nm}^{(-)}\right)^2}\right).
\end{equation}
\subsection{Comparison between HEOM and semiclassical dynamics}
Here, we compare the population dynamics obtained from HEOM simulations with those predicted by the semiclassical description. For the latter, we recall that the semiclassical master equation (ME) is given by
\begin{equation}
    \partial_t P_m (t) = \sum_{n}(\underbrace{\Gamma_{mn} P_{n}(t)}_{\rm ingoing\,\, flow} - \underbrace{\Gamma_{nm} P_{m}(t)}_{\rm outgoing \,\,flow}),
    \label{eq:semi_master}
\end{equation}
where the rate is given by 
\begin{equation}
\Gamma_{mn} = \frac{e^{-2|j_m-j_0|/\xi} e^{-2|j_n-j_0|/\xi}}{2\mathcal{N}^2}
\left( \frac{g\gamma^2}{\gamma^2 + \left((\epsilon_n - \epsilon_m) + \omega_0\right)^2} + \frac{ g\gamma^2 }{\gamma^2 + \left((\epsilon_n - \epsilon_m) - \omega_0\right)^2}\right).
\end{equation}
Under the assumption that $g = c\, \kappa$, i.e., that the coupling to the noise is proportional to the strength of the bath correlation function, we obtain
\begin{align}
&\Gamma_{mn} \sim \underbrace{e^{-2|j_m-j_0|/\xi} e^{-2|j_n-j_0|/\xi}}_{\textrm{exponential\, localization}} \underbrace{L(\epsilon_m - \epsilon_n)}_{\rm bath-induced\, activation}, \\& L(\epsilon_m - \epsilon_n)\equiv\frac{1}{2}\left[\frac{\kappa\gamma^2}{\gamma^2 + \left((\epsilon_n - \epsilon_m) + \omega_0\right)^2} + \frac{ \kappa\gamma^2 }{\gamma^2 + \left((\epsilon_n - \epsilon_m) - \omega_0\right)^2}\right].
\end{align}
For $\omega_0=0$, $L(\omega)$ reduces to a single Lorentzian and can be interpreted as the probability that the noise induces incoherent hopping processes in the system.
\begin{figure}[!t]
\centering
\includegraphics[width=0.68\linewidth]{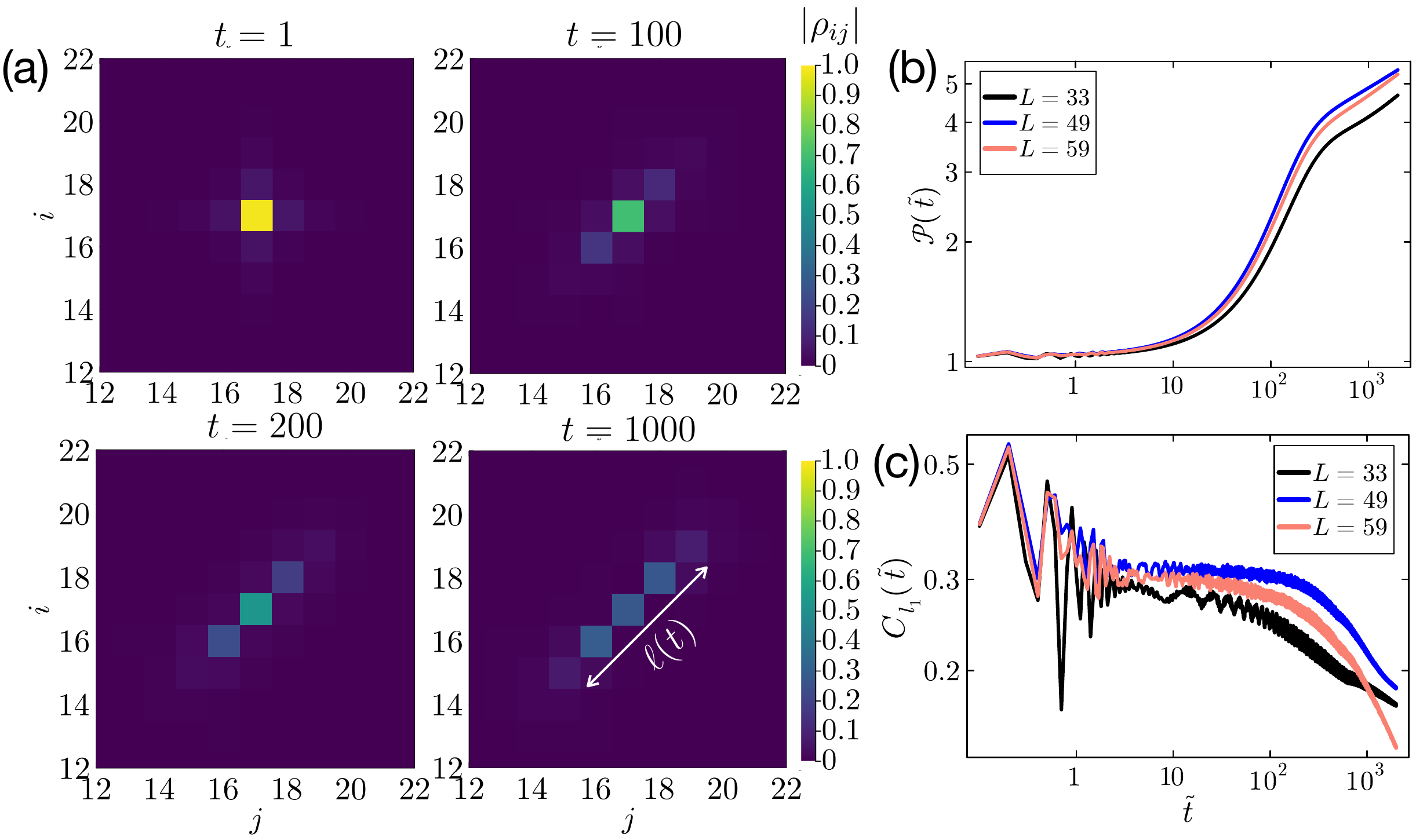}
\caption{Particle spreading along the chain in the deeply localized phase ($h=10$). Panel (a): density matrix elements $|\rho_{ij}(t)|$ in the neighborhood of the central site $j_0$ at different times $t$ for a system with $L=33$ sites in the deeply localized phase ($h=10$). Panel (b) and (c): spreading measure $\mathcal{P}(t)$ (b) and $l_1$-norm of coherence (c) for different system sizes $L$. The other parameters are set to $\kappa=2,\, \gamma=10, \, \omega_0=0, \textrm{tier}=6$. For this set of parameters $\Gamma_{\rm eff}=1$ and so $t=\tilde{t}$.}
\label{fig:spreading}
\end{figure}

We focus on the deeply localized regime, where $\ket{m} \approx \ket{j_m}$, implying that $P_n(t) \approx P_{j_n}(t)$. In this limit, the rates $\Gamma_{mn} \approx \Gamma_{j_m,j_n}$ can be interpreted as effective incoherent hopping rates. Moreover, the eigenenergies are well approximated by $\epsilon_n = h \cos(2\pi \beta j_n)$.

We aim to assess whether the semiclassical and quantum (HEOM) descriptions quantitatively agree in this regime. The semiclassical dynamics is initialized in the eigenstate localized at site $j_0$; the corresponding populations, denoted by $P_j^{(\rm CL)}$, are collected in the vector $\boldsymbol{P}^{(\rm CL)}(t)=\left[P_1^{(\rm CL)}(t), \ldots, P_L^{(\rm CL)}(t)\right]$, which in this limit closely approximates the on-site populations. In parallel, we extract the populations $\boldsymbol{P}(t)=\left[P_1(t), \ldots, P_L(t)\right]$ from HEOM and quantify the deviation between the two descriptions as
\begin{equation}
D_{\rm HEOM,CL}(t)=\frac{1}{2}\sum_j \left|P_j(t)-P_j^{(\rm CL)}(t)\right|.
\end{equation}
We then vary the proportionality constant $c$ to identify the regime of optimal agreement [Fig.~\ref{fig:comparison_heom_semi}], and analyze $D_{\rm HEOM,CL}(t)$ as a function of $\gamma$, $\kappa$, and $\omega_0$. Exploring different parameter sets, we find that the agreement is optimized for $c \approx 0.1$. Accordingly, in the following we fix $g = \kappa/10$ when comparing HEOM and semiclassical dynamics.
\subsection{RMSD with semiclassical dynamics}
We analyze the dynamics of the semiclassical ME in the deeply localized regime ($h=10$) with $\omega_0=0$ by initializing the system in the eigenstate localized around site $j_0$, which closely approximates the state $\ket{j_0}$. We then compute the semiclassical RMSD,
\begin{equation}
    \sigma_{\rm CL}(t)=\sqrt{\sum_{m=1}^{L}(m-j_0)^2 P_{m}^{(\rm CL)}(t)}.
\end{equation}
To facilitate comparison across different parameters, time is rescaled by the effective rate $\Gamma_{\rm eff}^{(\rm CL)}=\frac{10\,g\,\gamma^2}{\gamma^2 + h^2}$, which corresponds to $L(h)$ for $\omega_0=0$ and $g=\kappa/10$. This choice is motivated by the result of Fig.~\ref{fig:comparison_heom_semi} discussed in the previous section. Rescaled times are denoted by $\tilde{t}$. Moreover, since we consider the deeply localized regime, we approximate the eigenergies as $\epsilon_m=h\cos(2\pi\beta m)$.

\begin{figure}[!t]
\centering
\includegraphics[width=0.9\linewidth]{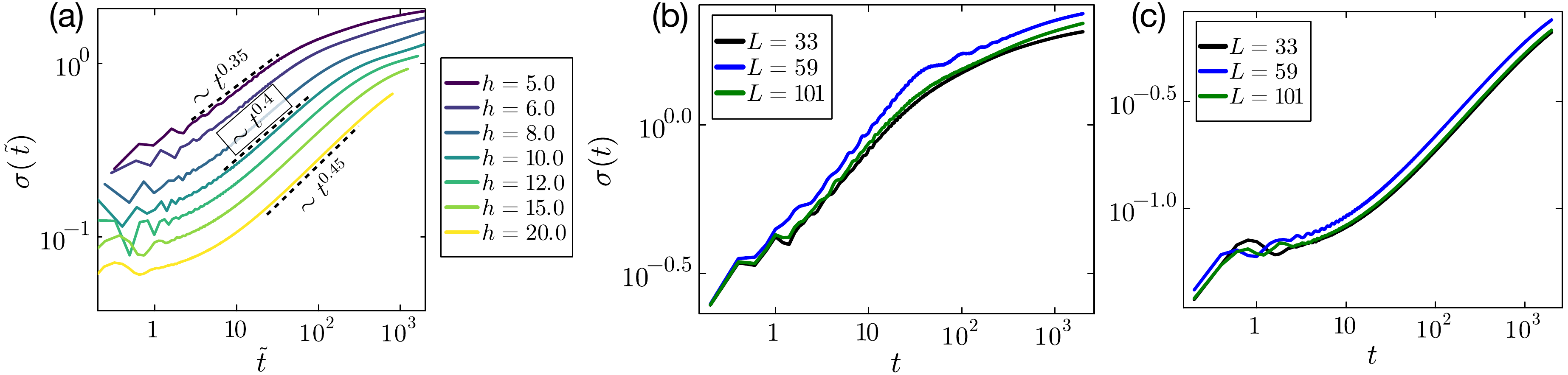}
\caption{Effect of the AAH field $h$ on the RMSD dynamics. Panel (a): RMSD dynamics for $L=33, \gamma=10, \kappa=2, \omega_0=0, \textrm{tier}=6$ and different values of $h$, times are rescaled by the factor $\Gamma_{\rm eff}$; the black dashed lines indicate power-law fits. Panels (b) and (c): $L=33, \gamma=10, \kappa=2, \omega_0=0, \textrm{tier}=6$, with $h=5$ (panel (a)) and $h=20$ (panel (b)).}
\label{fig:diffh}
\end{figure}
Fig.~\ref{fig:semicalssical_dyn}(a) shows the semiclassical dynamics for various arbitrary choices of the parameters $g$ (or $\kappa$) and $\gamma$. We observe that the time rescaling is effective, as the curves collapse onto a common behavior. Quantitatively, the semiclassical description captures an initial diffusive regime, $\sigma_{\rm CL}(\tilde{t}) \sim \tilde{t}^{1/2}$, followed, consistently with the HEOM results, by the emergence of a subdiffusive dynamics characterized by $\sigma_{\rm CL}(\tilde{t}) \sim \tilde{t}^{0.142}$. In Fig.~\ref{fig:semicalssical_dyn}(b) we compare the semiclassical and HEOM dynamics. The main differences between the two approaches arise at short times: in the HEOM evolution, the particle initially exhibits small fluctuations around the central site (the RMSD is weakly larger than zero) and only later begins to spread, as indicated by the slow growth of $\sigma_{\rm HEOM}$. In contrast, within the semiclassical description, the particle displays diffusive motion from the outset, before transitioning to a different dynamical regime characterized by subdiffusive behavior. We emphasize that, at long times ($\tilde{t} \gtrsim 10^2$), the two descriptions converge and yield the same transport behavior. In contrast, at short times, HEOM predicts subdiffusive dynamics, unlike the diffusive behavior predicted by the semiclassical ME.
\subsection{RMSD collapse in the deeply localized regime}
We discuss the collapse of the RMSD dynamics under the time rescaling $\tilde{t} = \Gamma_{\rm eff} t = \frac{\kappa \gamma^2}{\gamma^2 + h^2}\, t$. We consider a set of arbitrarily chosen dissipation parameters with $\omega_0 = 0$, and three different system sizes $L$; the results are shown in Fig.~\ref{fig:collapse}.

We observe a clear collapse of the RMSD curves, particularly at short times. This behavior indicates that, after an initial transient characterized by slow growth, transport follows the scaling form $\sigma(t) \approx D\, \tilde{t}^{\alpha} = \left[ D\, \Gamma_{\rm eff}^{\alpha} \right] t^{\alpha}$. Here, $D$ is a subdiffusive prefactor that is independent of the dissipation parameters $\gamma$ and $\kappa$, but depends on the localization properties of the AAH eigenmodes. The resulting ultraslow dynamics exhibits anomalous subdiffusion, with $\sigma(\tilde{t}) \sim \tilde{t}^{\,0.4}$ at early times and $\sigma(\tilde{t}) \sim \tilde{t}^{\,0.15}$ at later times. The crossover occurs at $\tilde{\tau}_{\rm slow} = \Gamma_{\rm eff}\,\tau_{\rm slow}$, marking the onset of the slow HEOM eigenvalues cluster. These scaling behaviors are robust against variations of the dissipation parameters and, provided that an eigenstate remains localized near the center of the chain, persist across different system sizes.

\subsection{Incoherent bath-activated spreading of the particle in the deeply localized regime}
In this section, we analyze particle spreading in the deeply localized phase, showing that the dynamics are predominantly incoherent. To this end, we introduce the following quantifiers:
\begin{equation}
\mathcal{P}(t)\coloneqq \frac{1}{\sum_{j=1}^L P_j(t)^2}, \quad
C_{l_1}(t)\coloneqq \sum_{\substack{i,j \ i \neq j}} |\rho_{ij}(t)|,
\end{equation}
where $\mathcal{P}(t)$ is the population participation ratio (PPR), which measures how widely the particle spreads. It interpolates between the fully localized case, $\rho = \ket{j}\bra{j}$, where $\mathcal{P} = 1$, and the fully delocalized (uniform population) case, $P_j = 1/L$ for all $j$, where $\mathcal{P} = L$. More generally, for a state of the form $\rho = \sum_{j \in \ell} \frac{1}{d_{\ell}} \ket{j}\bra{j}$, one has $\mathcal{P} = d_{\ell}$, where $d_{\ell}$ is the number of sites over which the particle is spread. We stress that $\mathcal{P}(t)$ depends only on populations and is therefore insensitive to quantum coherence. To assess whether transport is genuinely incoherent, we also consider the $l_1$-norm of coherence, $C_{l_1}(t)$, which provides a well-defined measure of coherence~\cite{baumgratz2014quantifying}.

\begin{figure}[!t]
\centering
\includegraphics[width=0.7\linewidth]{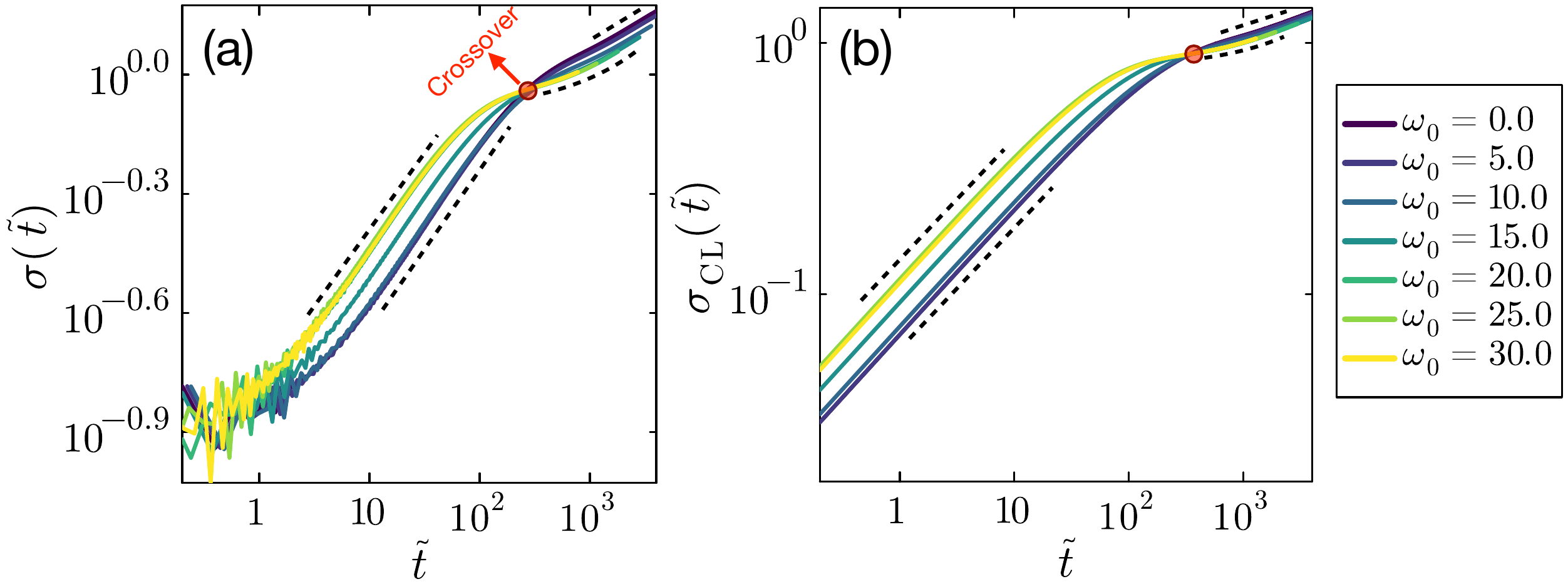}
\caption{Impact of $\omega_0$ on the RMSD dynamics in the deeply localized phase ($h=10$), computed with both HEOM and semiclassical ME. Panel (a) reports the HEOM results with $L=33, \gamma=10, \kappa=2, \textrm{tier}=6$. Panel (b) reports the semiclassical dynamics for $L=33, \gamma=10, g=\kappa/10$. Times are rescaled by repspectively $\Gamma_{\rm eff}$ and $\Gamma_{\rm eff}^{(\rm CL)}$.}
\label{fig:diffomega}
\end{figure}
Before analyzing these two quantifiers, we show in Fig.~\ref{fig:spreading}(a) the absolute value of the density matrix elements in the computational basis, $|\rho_{ij}|$, at four different times. We observe that a particle initially localized at the center of the chain starts to spread while maintaining low coherence. In particular, at the longest time considered ($t=10^3$), the diagonal elements are well spread over a region $\ell$, while coherences are strongly suppressed. The time-evolved state can thus be approximated as $\rho(t)\approx \sum_{j\in \ell(t)} P_j(t)\ket{j}\bra{j}$. This result confirms that the bath-activated hopping is predominantly incoherent.

In Fig.~\ref{fig:spreading}(b),(c), we show the dynamics of the PPR and the $l_1$-coherence norm for different system sizes $L$. We observe a behavior consistent with that of the RMSD. Regarding the PPR, after an initial slow escape from unity, the growth becomes steeper and then gradually slows down at longer times. Around the point where the growth changes slope, $\mathcal{P}(t)$ is approximately equal to four, indicating that the particle has spread to include the two nearest neighbors of the central site and slightly beyond. Regarding coherence, after pronounced oscillations at short times, we observe a plateau, consistent with the metastable regime suggested by the HEOM spectrum. At later times, the coherence measure decreases, indicating that incoherent dynamics takes over and the state approaches the diagonal form $\rho(t)\approx \sum_{j\in \ell(t)} P_j(t)\ket{j}\bra{j}$.
\section{Effect of the AAH field on the transport in the localized regime}
We investigate the influence of the AAH field $h$ on transport properties in the localized regime. In particular, we focus on the RMSD $\sigma(t)$ and its characteristic power-law scaling with time, $\sigma(t) \approx [D\, \Gamma_{\rm eff}^{\alpha}] t^\alpha$, which provides insight into the anomalous transport behavior induced by variations in the field strength.

In Fig.~\ref{fig:diffh}(a), we show the RMSD dynamics for different values of $h$ and find that the scaling exponent $\alpha$ increases with the field strength. This trend indicates that as $h$ grows, the transport gradually approaches a diffusive regime ($\alpha \to 1/2$). However, the overall magnitude of $\sigma(t)$ also depends on the prefactor $D$. When the time is rescaled as $\tilde{t} = \Gamma_{\rm eff} t$, the RMSD curves do not collapse, indicating that $D$ is itself $h$-dependent. In particular, $D$ decreases with increasing $h$, reflecting the stronger localization of eigenstates and the correspondingly slower transport dynamics. Therefore, while the exponent $\alpha$ grows with $h$, the absolute spreading of the RMSD is suppressed.

In Fig.~\ref{fig:diffh}(b) and (c), we show the RMSD dynamics for different system sizes and two values of $h$. These results demonstrate that, even for stronger fields (e.g., $h=10$ as shown in the main text), the qualitative behavior of the RMSD is largely independent of system size. To account for finite-size fluctuations, the data have been appropriately filtered, which removes spurious oscillations and highlights the underlying transport dynamics. This confirms that the observed scaling behavior of the RMSD is intrinsic to the localized regime and not an artifact of finite system size, supporting the robustness of our conclusions regarding the anomalous transport properties.
\section{Effect of $\omega_0$ on the transport in the deeply localized regime}
\begin{figure}[!t]
\centering
\includegraphics[width=\linewidth]{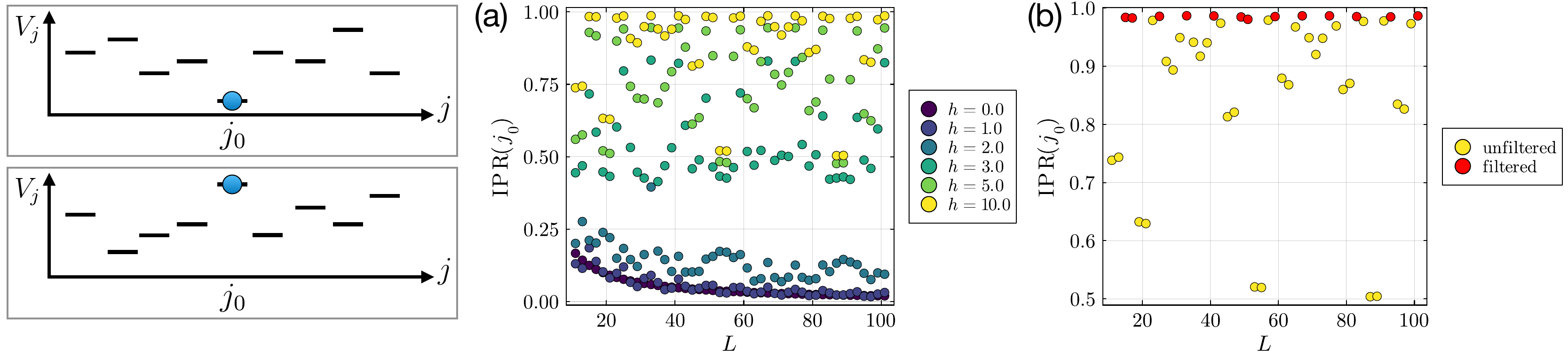}
\caption{The sketch on the left illustrates the ideal situation considered: the central site is located at a minimum or a maximum of the potential, so that the Hamiltonian admits an eigenstate that is well localized at the center. Panel (a): $\textrm{IPR}(j_0)$ in the AAH model as a function of the system size $L$ for different values of $h$. Panel (b): $\textrm{IPR}(j_0)$ as a function of the system size $L$ for $h=10$, in red we report the filtered system's sizes with $\textrm{IPR}(j_0)>1-\varepsilon$, with $\varepsilon=0.02$.}
\label{fig:iprfilter}
\end{figure}
%
%
We investigate the role of the dissipation parameter $\omega_0$ on the RMSD dynamics in the deeply localized regime (Fig.~\ref{fig:diffomega}). To perform time rescaling within the HEOM framework, we define the effective rate as
\begin{equation}
    \Gamma_{\rm eff} = L(h) = \frac{1}{2}\left[\frac{\kappa \gamma^2}{\gamma^2 + (h + \omega_0)^2} + \frac{\kappa \gamma^2}{\gamma^2 + (h - \omega_0)^2}\right].
\end{equation}
For the semiclassical simulations, we instead use
\begin{equation}
    \Gamma_{\rm eff}^{(\rm CL)} = \frac{1}{2}\left[\frac{10 g \gamma^2}{\gamma^2 + (h + \omega_0)^2} + \frac{10 g \gamma^2}{\gamma^2 + (h - \omega_0)^2}\right].
\end{equation}
Since in our case $g = \kappa / 10$, the two expressions coincide, yielding the same effective rate. We note that this rate is chosen following the same prescription as for $\omega_0 = 0$. In that limit, this choice led to a collapse of the RMSD in the semiclassical simulations (see Fig.~\ref{fig:semicalssical_dyn}) and to a good convergence onto a single curve in the HEOM results (see main text).

We now turn to the effect of $\omega_0$. We immediately observe that the same time rescaling no longer produces a collapse of the RMSD dynamics. Instead, two distinct behaviors emerge, which become clearly differentiated for $\omega_0 = 10$ and $\omega_0 = 20$ ($\omega_0/\gamma = 1$ and $2$), while $\omega_0 = 15$ exhibits an intermediate behavior.

For small $\omega_0$, both HEOM and semiclassical results retain the qualitative behavior observed at $\omega_0 = 0$: an initial subdiffusive regime in HEOM with $\sigma(t) \sim t^{0.4}$ (diffusive in the semiclassical case), followed by a slower, still subdiffusive growth. Conversely, for large $\omega_0$, the HEOM dynamics remains subdiffusive ($\sigma(t) \sim t^{0.4}$, diffusive semiclassically) but evolves faster in rescaled time, corresponding to a larger effective coefficient $D$. At later times, a crossover occurs (indicated by the red point in the figure), after which the spreading for small $\omega_0$ exceeds that for large $\omega_0$.

Beyond the crossover, the dynamics for large $\omega_0$ deviates from a power-law, as evidenced by its non-linear appearance on a log--log scale. This non-polynomial behavior is more pronounced in the full HEOM simulations. Overall, while increasing $\omega_0$ initially accelerates spreading in rescaled time, it ultimately slows the long-time dynamics. Therefore, $\omega_0$ affects the RMSD both qualitatively and quantitatively.
\section{Filtering the system's sizes}
In the localized phase, the eigenstates of the AAH model are exponentially localized in the thermodynamic limit. However, for finite system sizes and with a fixed phase $\phi=0$, exact exponential localization is not guaranteed. To address this, we focus on specific system sizes by filtering those for which the system possesses an eigenstate that is approximately localized at the site $j_0$. The ideal situation we consider is sketched in ~\ref{fig:iprfilter}. 

As a quantifier, we use the inverse participation ratio (IPR) of the state $\ket{j_0}$ in the basis of the AAH eigenstates:
\begin{equation}
    \textrm{IPR}(j_0) = \sum_m |\braket{m|j_0}|^4.
\end{equation}
If there exists an eigenstate that is approximately localized at $j_0$, then $\textrm{IPR}(j_0) \approx 1$. In Fig.~\ref{fig:iprfilter}(a), we report $\textrm{IPR}(j_0)$ as a function of $L$ for different values of the field $h$. This allows us to clearly distinguish between the extended ($h<2$) and localized ($h>2$) phases. While for large $h$ in the localized phase $\textrm{IPR}(j_0)$ is significantly larger than in the extended phase, we observe that for certain values of $L$ the IPR is noticeably smaller than one, indicating that no eigenstate is strongly localized at $j_0$.

For this reason, we filter out these system sizes and analyze only those with size $L$ for which there exists an eigenstate localized at $j_0$. Quantitatively, this corresponds to selecting only sizes for which $\textrm{IPR}(j_0) > 1-\varepsilon$, with $\varepsilon \ll 1$. In Fig.~\ref{fig:iprfilter}(b), we show the filtered system sizes using $\varepsilon = 0.02$. Specifically, we consider odd $L \leq 101$ and obtain the set of filtered sizes $\boldsymbol{L}_{\rm filtered} = [15,17,25,33,41,49,51,59,67,75,83,93,101]$. The choice of $\varepsilon$ is justified by the fact that these filtered sizes exhibit a similar dynamical behavior.

\section{Comparison of the Spectra}
\begin{figure}
    \centering
\includegraphics[width=\linewidth]{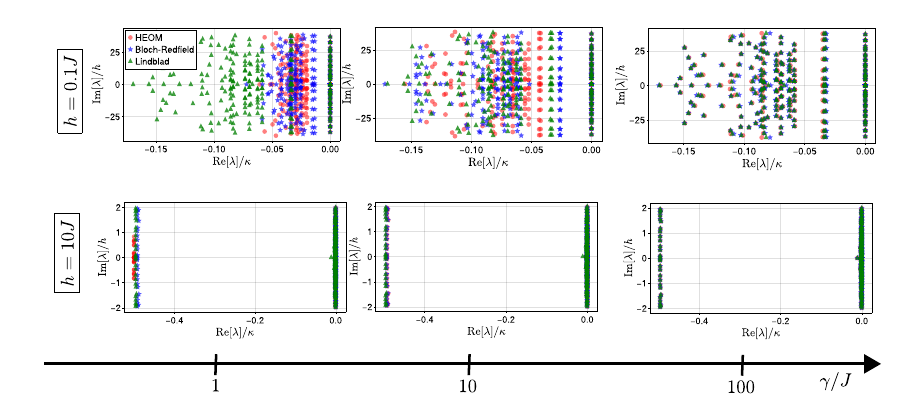}
    \caption{Comparison of the 225 eigenvalues with the largest real part associated with the HEOM generators (red circles), the Bloch-Redfield generator (blue stars) and the pure Lindblad limit (green triangle). For both the localized ($h=10J$) and the extended phase ($h=0.1J$), $\gamma/J \in \{1, 10, 100 \}$ in units of $J$. Parameters for all plots: $L = 15, \kappa = 2J$, $\omega_0 = 0 J$ and the tier is $5$. }
    \label{fig:spectra_comparison}
\end{figure}
In this section, we compare the spectra of the dynamical generators obtained using three different approaches: the Lindblad master equation, the Bloch-Redfield equation, and the Hierarchical Equations of Motion (HEOM). In all cases, the bath correlation function is taken to be exponentially decaying,
\begin{equation}
C(t) = \frac{\gamma \kappa}{2} e^{-\gamma |t|},
\end{equation}
which introduces a finite memory timescale controlled by $\gamma^{-1}$. While HEOM naturally incorporates non-Markovian effects, both the Lindblad and Bloch-Redfield approaches rely on the Markovian approximation and therefore neglect memory effects to different extents.

In Fig.~\ref{fig:spectra_comparison} we compare the spectra of the corresponding dynamical generators in two asymptotic regimes: the extended and localized phases. We begin by analyzing the extended phase. For small values of $\gamma$, the bath correlation time is long and the system lies deep in the non-Markovian regime. As a consequence, the spectral structure exhibits significant discrepancies between the three methods, with HEOM capturing features that are absent or only partially reproduced by the Markovian approaches. As $\gamma$ increases, the bath memory time decreases and the system progressively approaches the pure-Lindblad limit. In this regime, the spectra obtained from HEOM and Bloch-Redfield gradually converge towards that of the Lindblad generator. In the asymptotic limit $\gamma \gg J$, where the bath correlation time becomes negligible compared to the system timescale, all three spectra essentially coincide, indicating that memory effects no longer play a relevant role.

This behavior provides a clear interpretation of the qualitative differences observed in the system dynamics at different values of $\gamma$. In particular, the HEOM spectrum undergoes substantial structural changes as the $\gamma$ parameter is varied. These spectral transformations are directly connected to the distinct dynamical features observed across the different regimes.


Conversely, in the deeply localized regime, the spectra obtained from HEOM, Bloch-Redfield, and Lindblad-type descriptions display the same characteristic clustered organization already for small values of $\gamma$. As $\gamma$ increases and memory effects are progressively suppressed, the spectra tend to align, and the differences between the methods become negligible.  This behavior is consistent with the corresponding dynamical results, which indicate that in the localized phase the primary effect of the environment is just a renormalization of the relaxation timescales. In other words, dissipation effects predominantly manifest as a rescaling of time, rather than inducing qualitative changes in the structure of the dynamics. This is directly mirrored in the spectral properties, where the overall structure remains robust while the position of the eigenvalues inside the clusters may slightly shift.

\section{Proof that $\Delta_\mathrm{HEOM} \rightarrow \kappa/2$ for $h \rightarrow +\infty$.}
Let us demonstrate that in the limit $h \rightarrow +\infty$, $\mathrm{\Delta_\mathrm{HEOM}}$ tends to $\kappa/2$. For large $h$, the eigenstates of the Aubry-André Hamiltonian are highly localized since $\xi = 1/\mathrm{ln}(h/2J)$; in the formal limit $h\rightarrow +\infty$, there are perfectly localized. The Hamiltonian can then be written simply as 
\begin{equation}
    \hat{H}_f = \sum_{i} \epsilon_i \ket{i} \bra{i},
\end{equation}
while the coupling operator (i.e., the operator mediating the interaction between the system and the bath) reduces to $\hat{n}_{j_0} =  \ket{j_0} \bra{j_0}$ in the one-particle sector. As discussed in the previous section, we also know that for large $h$, the eigenvalues with the largest real part of the HEOM spectrum coincides with the ones of reduces approaches such as Bloch-Redfield or Lindblad. Consequently, we only need to prove that the spectrum of the Lindbladian 
\begin{equation}
    \mathcal{L} [\cdot] = -i [\hat{H}_f, \cdot] - \kappa \left(\hat{n}_{j_0} \cdot \hat{n}_{j_0} - \frac{1}{2} \{ \hat{n}_{j_0}, \cdot \}\right)
\end{equation}
is made of two clusters of eigenvalues separated by $\Delta = \kappa/2$. This is easily done by noting that $\ket{i} \bra{j}$ with both $i$ and $j$ different from $j_0$ or both of them equal to $j_0$ give 
\begin{equation}
    \mathcal{L}[\ket{i} \bra{j}] = -i (\epsilon_i - \epsilon_j) \ket{i} \bra{j}.
\end{equation}
These eigenoperators are then associated with the eigenvalues with zero real part, while $\ket{i} \bra{j}$ with either $i$ or $j$ equal to $j_0$ give 
\begin{equation}
    \mathcal{L}[\ket{i} \bra{j}] = -\kappa/2 -i (\epsilon_i - \epsilon_j) \ket{i} \bra{j}.
\end{equation}
Therefore, in this limit, the spectrum is made of bands separated by $\kappa/2$, and since the relevant HEOM eigenvalues converge to the Lindblad ones in this limit, the HEOM gap tends to $\kappa/2$.  
\end{document}